\documentclass[12pt,preprint]{aastex}
\usepackage{natbib}
\usepackage{amsmath}
\usepackage{comment}
\usepackage{color}

\newcommand{\bea}{\begin{eqnarray}}
\newcommand{\eea}{\end{eqnarray}}
\def\beq{\begin{equation}}
\def\eeq{\end{equation}}

\def\f{{\it Fermi}\ }

\def\he3{$^3$He\,}
\def\he4{$^4$He\,}

\begin{document}
\title{Probing the Puzzle of Behind-the-Limb $\gamma$-ray Flares: Data-driven Simulations of Magnetic Connectivity and CME-driven Shock Evolution}
\author{Meng Jin\altaffilmark{1,2}, Vahe Petrosian\altaffilmark{3,4}, Wei Liu\altaffilmark{1,5}, Nariaki V. Nitta\altaffilmark{1}, Nicola Omodei\altaffilmark{3}, Fatima Rubio da Costa\altaffilmark{3}, Frederic Effenberger\altaffilmark{5,6}, Gang Li\altaffilmark{7}, Melissa Pesce-Rollins\altaffilmark{8}, Alice Allafort\altaffilmark{3}, \& Ward Manchester IV\altaffilmark{9}}

\altaffiltext{1}{Lockheed Martin Solar and Astrophysics Lab, Palo Alto, CA 94304, USA; jinmeng@lmsal.com}
\altaffiltext{2}{SETI Institute, Mountain View, CA 94043, USA}
\altaffiltext{3}{Department of Physics, Stanford University, Stanford, CA 94305, USA}
\altaffiltext{4}{Department of Applied Physics, Stanford University, Stanford, CA 94305, USA}
\altaffiltext{5}{Bay Area Environmental Research Institute, NASA Research Park, Moffett Field, CA 94035, USA}
\altaffiltext{6}{Helmholtz Centre Potsdam, GFZ, German Research Centre for Geosciences, Potsdam, Germany}
\altaffiltext{7}{University of Alabama in Huntsville, Huntsville, AL 35899, USA}
\altaffiltext{8}{Istituto Nazionale di Fisica Nucleare, Sezione di Pisa, I-56127 Pisa, Italy}
\altaffiltext{9}{Department of Climate and Space Sciences and Engineering, University of Michigan, Ann Arbor, MI 48109, USA}

\begin{abstract}
Recent detections of high-energy $\gamma$-rays from behind-the-limb (BTL) solar flares by the \emph{Fermi $\gamma$-ray Space Telescope} pose a puzzle and challenge on the particle acceleration and transport mechanisms. In such events, the $\gamma$-ray emission region is located away from the BTL flare site by up to tens of degrees in heliogrpahic longitude. It is thus hypothesized that particles are accelerated at the shock driven by the coronal mass ejection (CME) and then travel from the shock downstream back to the front side of the Sun to produce the observed $\gamma$-rays. To test this scenario, we performed data-driven, global magnetohydrodynamics simulations of the CME associated with a well-observed BTL flare on 2014 September 1. We found that part of the CME-driven shock develops magnetic connectivity with the $\gamma$-ray emission region, facilitating transport of particles back to the Sun. Moreover, the observed increase in $\gamma$-ray flux is temporally correlated with (1) the increase of the shock compression ratio and (2) the presence of a quasi-perpendicular shock over the area that is magnetically connected to the $\gamma$-ray emitting region, both conditions favoring the diffusive shock acceleration (DSA) of particles. These results support the above hypothesis and can help resolve another puzzle, i.e., long-duration (up to 20 hours) $\gamma$-rays flares. We suggest that, in addition to DSA, stochastic acceleration by plasma turbulence may also play a role, especially in the shock downstream region and during the early stage when the shock Alfv\'{e}n Mach number is small. 
\end{abstract}


\section{Introduction}
\label{sect_intro}

The Large Area Telescope (LAT) on board the \emph{Fermi $\gamma$-ray Space Telescope} (\emph{Fermi}: \citealt{atwood09}) has increased the number of observed solar flares with photon emission above 100 MeV by an order of magnitude compared to all previous instruments \citep{share17}. One prominent characteristic of these flares is the long-duration emission extending hours past the impulsive phase, long after other flare associated electromagnetic emissions at longer wavelengths have decayed (e.g., \citealt{ajello14}). High-energy $\gamma$-rays can be produced by electrons and ions (primarily protons), with somewhat larger energies than the photons, via relativistic electron bremsstrahlung or decay of pions (and their byproducts) produced by interactions of protons with the background ions. Both mechanisms require transport of the accelerated particles deep into the high-density solar photosphere (where the particles lose most of their energy) and  through column depths of about $2.5\times 10^{25}$ and $2.5\times 10^{26}$~cm$^{-2}$, respectively. Thus, the accelerated particles, wherever produced, must travel to the photosphere to produce the observed $\gamma$-rays. This transport is guided by the magnetic field lines connecting the acceleration site to the photosphere. 

The impulsive phase (duration of $< 10^3$ s) radiations (from microwaves to GeV $\gamma$-rays) are produced by the interactions of the nonthermal electrons and protons with the flaring loop magnetic field and plasma (mainly at the loop footpoints). In a vast majority of flares, the impulsive emission is dominated by nonthermal electrons, rather than protons \citep{shih09}, in which particles are generally believed to be accelerated near the loop-top region (heights $> 10 ^9$ cm; e.g., \citealt{petrosian02, liu13}). This acceleration mechanism may be at work for some of the long-duration $\gamma$-ray flares observed by \f (e.g., \citealt{ajello14}). Examples include the 2011 March 7\,--\,8 \citep{ackermann14} and the 2017 September 10 \citep{omodei18} flares, during which the centroid of the $\gamma$-ray source coincided well with the active region (AR) where the flare was initiated. This is also the case for the initial phase of the stronger 2012 March 7 flare, which lasted for about 20 hours, later with a temporal drift of the centroid away from the AR \citep{ajello14}. In these events, the bulk of the hard X-rays (HXRs) and $\gamma$-rays may be explained in terms of thick-target emission from the footpoints of the flaring loops.

However, \f also detected $>$100~MeV photons from three other flares which, according to observations by the \emph{Solar TErrestrial RElations Observatory} (\emph{STEREO}), originated from ARs that were located 13$\arcdeg$\,--\,36$\arcdeg$ behind the solar limb seen from the Earth perspective \citep{pesce15, ackermann17}. These flares were also detected in HXRs by the \emph{Reuven Ramaty High Energy Solar Spectroscopic Imager}  (\emph{RHESSI}), \emph{Fermi}/Gamma-ray Burst Monitor (GBM), and \emph{Wind}/Konus with similar time profiles, in extreme ultraviolet (EUV) by \emph{Solar Dynamics Observatory} (\emph{SDO}) and \emph{STEREO}, and in microwave by the Radio Solar Telescope Network (RSTN). In addition, \emph{RHESSI} detected HXR emission located just over the limb which is consistent with the top of the (relatively tall) flaring loop rooted at the source AR behind the solar limb. \emph{An important question is whether or not the LAT $\gamma$-rays are coming from this loop-top source as well}. As described below, the LAT observations and some theoretical arguments lead us to consider a different location on the Sun for the $>$$100$~MeV $\gamma$-ray source and perhaps a different site and mechanism for acceleration of particles (either electrons or protons). This would be an important step toward resolving the puzzle of \f BTL flares and understanding $\gamma$-ray flares in general.

\citet{cliver93} first proposed that the BTL $\gamma$-ray events are caused by particles that are accelerated at the shock driven by the associated coronal mass ejection (CME) and then propagate back to the visible solar disk. The goal of this study is to explore this scenario by investigating the magnetic connectivity and evolution of the CME-driven shock, and their relationship, in both space and time, with the observed $\gamma$-ray emission during a BTL flare. Specifically, we will evaluate to what extent the CME and CME-driven shock are magnetically connected to $\gamma$-ray emitting areas of the visible disk away from the AR, and track several key shock parameters over those magnetically-connected areas of the shock surface.\footnote{A corollary of this scenario is that we would expect a similar spread of $\gamma$-ray emission over the solar disk for on-disk flares as well. In fact, the new PASS-8 analysis of the X5.4 flare on 2012 March 7 shows hints of migration of the emission centroid moving away from its host AR over time (Allafort et al., in preparation), although in general it is harder to distinguish between a point source (which was assumed in locating the centroids) and an extended one due to the relative low number of photons detected by \emph{Fermi}.} To this end, we performed high-fidelity, data-driven magnetohydrodnamic (MHD) simulations to reconstruct the global corona and solar wind environment for the CME eruption associated with the strongest of the three BTL flares: {\it SOL2014-09-01}. We note the pioneering work by \citet{plotnikov17} toward this direction that used a potential magnetic field and a \emph{static} MHD global corona solution. We believe that the inclusion of the dynamic evolution of the CME, as done in the present study, is an important step forward and can shed critical new light on the underlying physics of BLT $\gamma$-ray flares.

This article is organized as follows. In Section~\ref{sect_obs}, we present a summary of relevant observations of this event and theoretical arguments. In Section~\ref{sect_modeling}, we describe our numerical model and present the simulation results with a focus on the magnetic field connectivity and shock evolution, followed by discussions in Section~\ref{sect_discuss} and summary and conclusion in Section~\ref{sect_summary}.

\section{Review of Observations and Theoretical Arguments: the {\it SOL2014-09-01} Flare \& CME Event}
\label{sect_obs}

Here we briefly review the observations of the \emph{SOL2014-09-01} (hereafter Sept14) flare relevant to this work (We refer the reader to \citealt{ackermann17} for more details). Specifically, we give two empirical reasons why we favor the CME-shock origin, rather than the direct flare acceleration, of the particles responsible for the $\gamma$-rays detected by the LAT in this flare.

The first reason is the difficulty of producing strong $\gamma$-rays in the tenuous solar corona. According to \emph{STEREO-B} data, this flare originated from what was named NOAA AR~12158 later. It was located at N14E126, about $36^{\circ}$ behind the east solar limb. \emph{RHESSI} images show a HXR source with a size of about 40$\arcsec$ ($\sim 30$ Mm) just over the limb, which is consistent with (a part of) the loop-top source of a relatively large flaring loop with a height of $\gtrsim$130~Mm above the photosphere. Similar examples have been reported \citep{krucker07}. Since all other $<$100~MeV emissions, as seen by \emph{Fermi}/GBM and \emph{Wind}/Konus, have light curves very similar to the \emph{RHESSI} HXRs, it is reasonable to assume that they also come from the top of the flare loop through thin-target bremsstrahlung emission \citep{chen13, petrosian16, effenberger17}. Since only a small fraction of the particle energy is lost during the {\it thin-target} bremsstrahlung, coronal HXR/$\gamma$-ray emission in general requires a higher number (and energy) of accelerated particles than if we are dealing with a {\it thick-target} footpoint emission, where particles lose all their energy \citep{petrosian73}, and where most of the HXRs and $\gamma$-rays \citep[e.g.,][]{hurford03,hurford06} in on-disk flares originate from. This difference would also be the case for either electrons or protons if the \emph{Fermi}/LAT $\gamma$-rays were also coming from the thin-target loop-top source. However, assuming thick-target emission by protons in the photosphere, one requires a total energy in protons comparable to that calculated for disk flares. Therefore, if the Sept14 \emph{Fermi}/LAT emission were from the loop-top source, it would require a much higher energy of the accelerated protons than any of the other (even stronger) \emph{Fermi}/LAT flares \citep{petrosian18}. This difficulty is the first reason for considering a different source and possibly a different acceleration mechanism for the production of the $\gamma$-rays.

The second and more important reason is that the centroid of the \emph{Fermi}/LAT source is about $300\arcsec (\sim 200$ Mm) northwest of the \emph{RHESSI} source and the corresponding light curve is different than that of all the other emissions. Specifically, the LAT light curve decays very gradually with emission detected for almost two hours, while all other emissions last less than one hour before falling below background. Although the possibility of LAT emission being also produced (in part) by particles accelerated near the loop-top source cannot be completely ruled out, the above two reasons lead to a more plausible scenario that this emission is produced at the photosphere by particles (most likely protons) accelerated at the CME-driven shock and escaping from the downstream back to the Sun \citep{cliver93}. In the case of BTL flares, unlike in on-disk flares, these particles must be streaming down to the photosphere along field lines connected to the LAT centroid region located on the visible disk, tens of degrees away from the host AR.

This scenario is further supported by the facts that the Sept14 flare is also associated with: (i) a fast CME observed by both \emph{SOHO}/LASCO and \emph{STEREO-B}/COR1 with a speed $>$1900~km~s$^{-1}$; (ii) a Type II radio burst with an estimated velocity of 2079 km s$^{-1}$ \citep{pesce15}, and (iii) an SEP event with a quick onset and hard spectrum observed by \emph{STEREO} \citep{cohen16, zelina17}. The CME white-light images and height-time history are shown in Figures~\ref{fig:wl} and \ref{fig:htplot}, respectively, and will be further discussed in \S 3.1.

\section{Modeling the {\it SOL2014-09-01} Event}
\label{sect_modeling}

\subsection{Global Coronal \& CME Models}
\label{subsect_model-descr}

To reconstruct the global corona and solar wind environment during the {\it SOL2014-09-01} CME eruption, we used the University of Michigan Alfv\'{e}n Wave Solar Model (AWSoM; \citealt{sokolov13, bart14}) within the Space Weather Modeling Framework (SWMF; \citealt{toth12}). AWSoM is a data-driven global MHD model with the inner boundary specified by observed magnetic maps and the simulation domain extending from the upper chromosphere to the corona and heliosphere. Physical processes implemented in the model include multi-species thermodynamics, electron heat conduction (both collisional and collisionless formulations), optically thin radiative cooling, and Alfv\'{e}n-wave turbulence that energizes the solar wind plasma. The Alfv\'{e}n-wave description is physically self-consistent, including non-Wentzel-Kramers-Brillouin (WKB) reflection \citep{heinemann80, velli93, hollweg07} and physics-based apportioning of turbulence dissipative heating to both electrons and protons. AWSoM has demonstrated its capability of reproducing solar corona condition with high-fidelity \citep[e.g.,][]{sokolov13, bart14, oran13, oran15, jin16, jin17a}.

Based on the steady-state global corona and solar wind solution, we initiate the CME by using an analytical Gibson-Low (GL) flux-rope model \citep{gibson98}, which has been successfully used in numerous modeling studies of CMEs (e.g., \citealt{chip04a, chip04b, lugaz05, chip14, jin16, jin17a}). The GL flux rope is mainly controlled by five parameters: the stretching parameter $a$ determines its shape, the distance $r_{1}$ of the flux rope center from the center of the Sun determines its initial position before being stretched, the radius $r_{0}$ of the flux-rope torus determines its size, $a_1$ determines its magnetic field strength, and a helicity parameter determines its positive (dextral) or negative (sinistral) helicity. Analytical profiles of the GL flux rope are obtained by finding a solution to the magnetohydrostatic equation $(\nabla\times{\bf B})\times{\bf B}-\nabla p-\rho {\bf g}=0$ and the solenoidal condition $\nabla\cdot{\bf B}=0$. This solution is derived by applying a mathematical stretching transformation $r\rightarrow r-a$ to an axisymmetric, spherical ball of twisted magnetic flux with radius $r_0$ centered in the heliospheric coordinate system at $r=r_1$. The transformed flux rope appears as a tear-drop shape. At the same time, Lorentz forces are introduced, which lead to a density-depleted cavity in the upper portion and a dense core at the lower portion of the flux rope, corresponding to a coronal cavity and a dense prominenence, respectively. This configuration can thus readily reproduce the typical three-part structure of an observed CME \citep{llling85}. The GL flux rope and contained plasma are then superposed onto the steady-state AWSoM solution of the solar corona: i.e. $\rho=\rho_{0}+\rho_{\rm GL}$, ${\bf B = B_{0}+B_{\rm GL}}$, $p=p_{0}+p_{\rm GL}$. The temperature will be updated from the new density $\rho$ and pressure $p$. The resulting combined background-flux rope system is in a state of force imbalance, due to the insufficient background plasma pressure to counter the magnetic pressure of the flux rope, and thus erupts immediately when the numerical model advances in time.

\begin{figure}[h]
\begin{center}$
\begin{array}{c}
\includegraphics[scale=0.8]{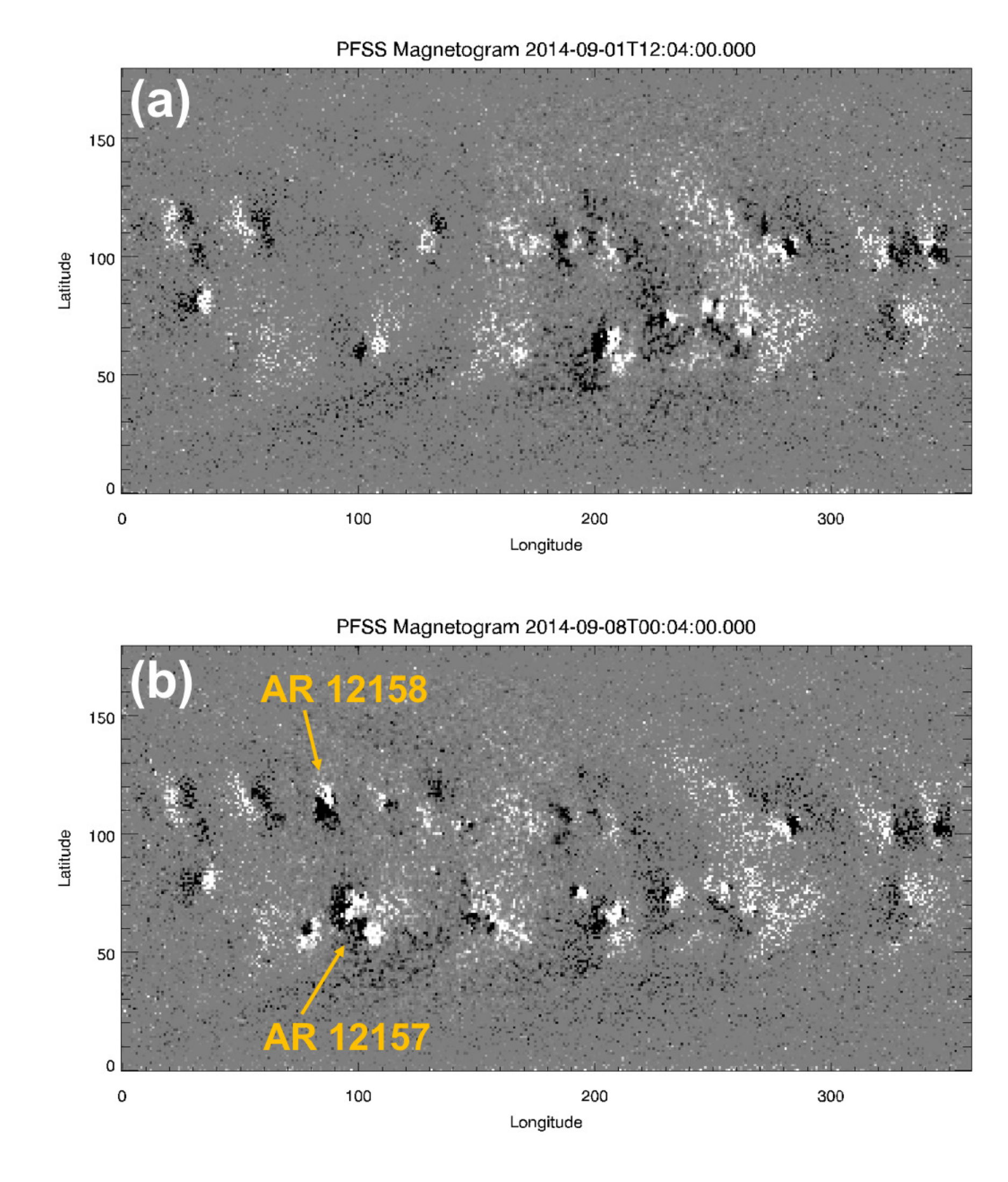}
\end{array}$
\end{center}
\caption{Synchronous magnetograms at (a) 2014-09-01 12:04:00 UT during the Sept14 \emph{Fermi} BTL $\gamma$-ray flare and (b) 2014-09-08 00:04:00 UT, one week after the flare, which has incorporated the flare hosting AR~12158 and is used to reconstruct the initial global coronal magnetic field in this study.}
\label{fig:magnetogram}
\end{figure}

To specify the inner boundary condition of the magnetic field, we utilize a global magnetic map sampled from an evolving photospheric flux transport model \citep{schrijver03}, which assimilates new observations within 60$^{\circ}$ from disk center obtained by the \emph{SDO} Helioseismic and Magnetic Imager \citep[HMI;][]{schou12}. The assimilated magnetogram is updated every 6 hours. The Sept14 flare occurred behind the east limb where no direct observation of the magnetic field is available. This means that the magnetic field around the flare site at the time of the event contains the most aged observation obtained from about a half solar rotation earlier when the region was on the western side of the visible solar disk. Therefore, a large amount of magnetic flux could potentially be missing. From the magnetogram closer in time shown in Figure \ref{fig:magnetogram}a, we find that the flare source region AR~12158 is indeed completely missing. To alleviate this problem, we choose the assimilated magnetogram on 2014 September 8 00:04:00 UT (Figure  \ref{fig:magnetogram}b), about a week after the event on September 1, when the magnetic field around the source region was first assimilated into the flux transport model. The missing flare source region AR~12158 and another large AR~12157 to the south of it are now properly included. The rest of the old and new magnetic maps are qualitatively very similar. As such, the 2014 September 8 magnetogram is a reasonable representation of the photospheric magnetic field at the time of the Sept14 flare and is thus used to specify the inner boundary condition of our global magnetic field model.

\begin{figure}[h]
\begin{center}$
\begin{array}{c}
\includegraphics[scale=1.1]{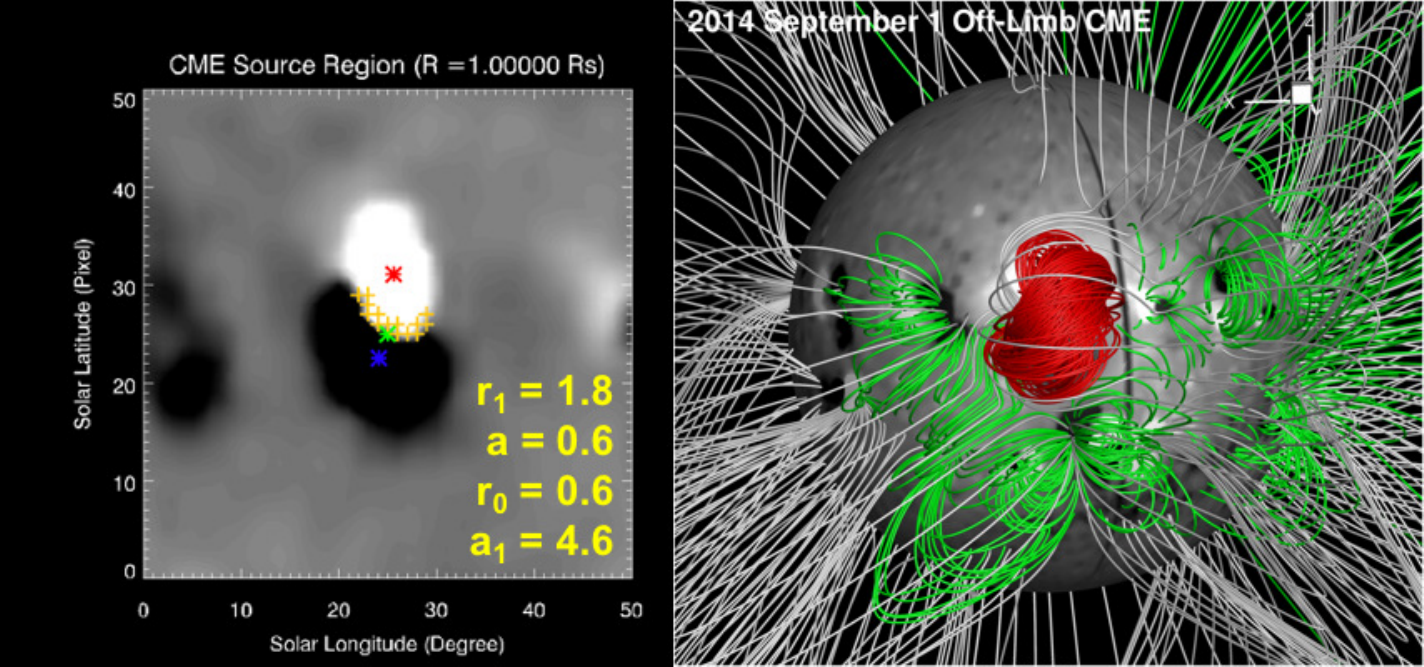}
\end{array}$
\end{center}
\caption{Left Panel: Zoom-in magnetic map of AR 12158. The red and blue symbols represent the weighted centers of positive and negative polarities. The yellow symbols represent the polarity inversion line. The green symbol shows the location for inserting the GL flux rope whose key control parameters are listed in the lower-right corner. Right Panel: The 3D initial configuration of the solar corona with the GL flux rope. The red field lines represent the initial flux rope. The white field lines represent the large-scale helmet streamers. The green field lines are selected surrounding active region as well as open field lines.}
\label{fig:initiation}
\end{figure}

To configure a proper GL flux rope for initiating the Sept14 CME, we utilize a newly developed tool called the Eruptive Event Generator using Gibson-Low configuration (EEGGL; \citealt{jin17b}), which is designed to determine the GL flux-rope, including its location, orientation, and five key controlling parameters, using the observed magnetogram and CME speed near the Sun. The left panel of Figure~\ref{fig:initiation}a shows a zoom-in view of AR 12158 with weighted centers of positive/negative polarities and the polarity inversion line (PIL) determined by EEGGL. The green asterisk marks the central location to insert the GL flux rope, whose calculated key parameters are also listed. The right panel shows the 3D configuration of the global coronal magnetic field, with the inserted GL flux rope shown in red. The white field lines represent the large-scale helmet streamer structures. The selected field lines from surrounding active regions and open field are marked in green. The GL flux rope erupts due to the force imbalance upon insertion into the active region. The simulation is then evolved forward in time and the MHD equations are solved in conservative forms to guarantee the energy conservation across the CME-driven shock \citep{bart10, chip12, jin13}. To better resolve the shock structure, two more levels of refinement along the CME path are performed, which make the cell size $\sim$0.02 R$_\odot$ at 2 R$_\odot$ and $\sim$0.06 R$_\odot$ at 5 R$_\odot$. We run the simulation for 1 hour after the initiation, until the CME reaches $\sim$10 R$_\odot$.

\begin{figure}[h]
\begin{center}$
\begin{array}{c}
\includegraphics[scale=0.85]{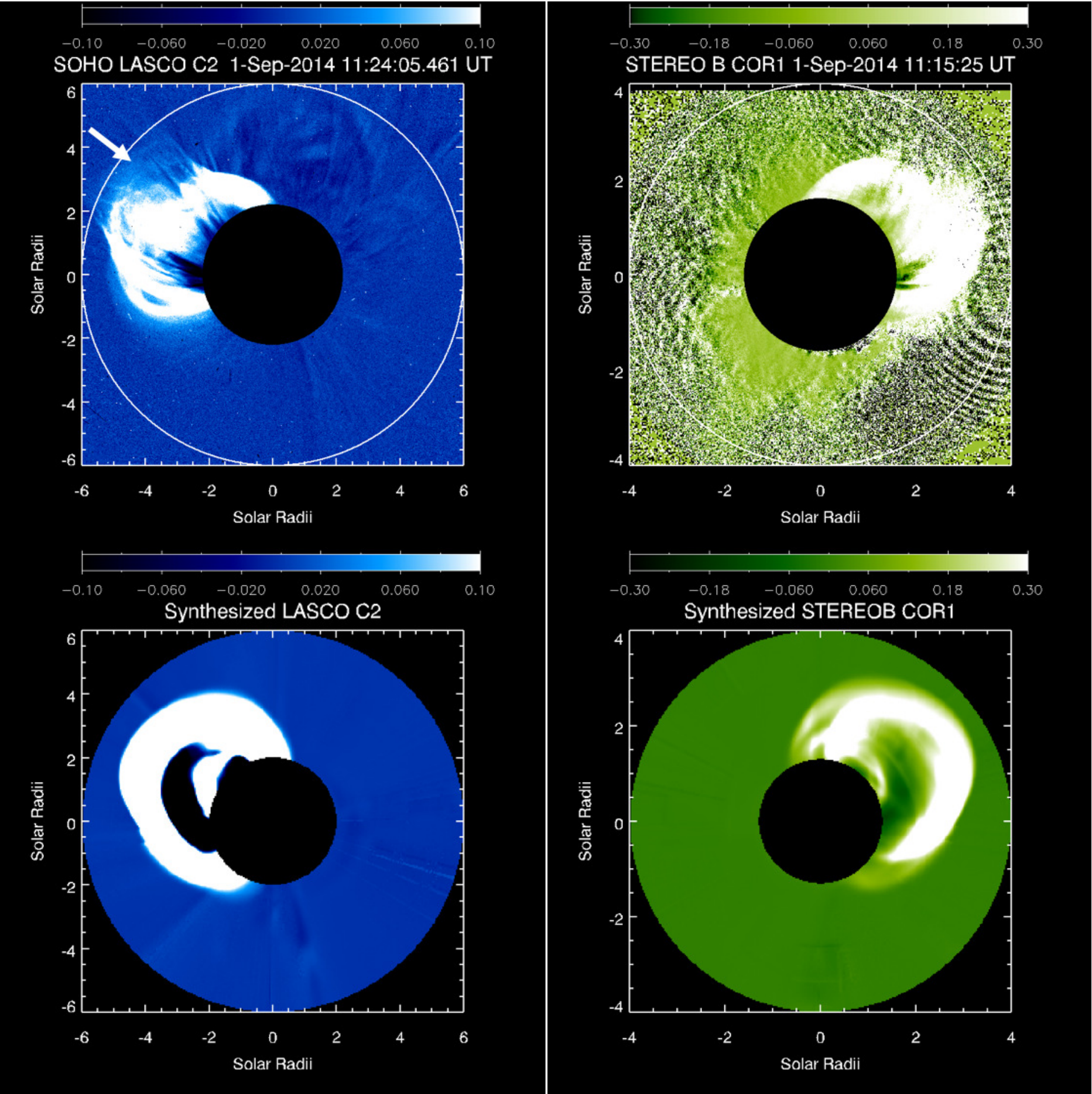}
\end{array}$
\end{center}
\caption{Comparison showing a general agreement between the white-light observations from \emph{SOHO} LASCO C2 (top left) and \emph{STEREO-B} COR1 (top right) and the respective synthesized white-light images from the simulation (bottom). The color scale shows the relative total brightness changes compared to the pre-event background level.}
\label{fig:wl}
\end{figure}

\begin{figure}[h]
\begin{center}$
\begin{array}{c}
\includegraphics[scale=0.3]{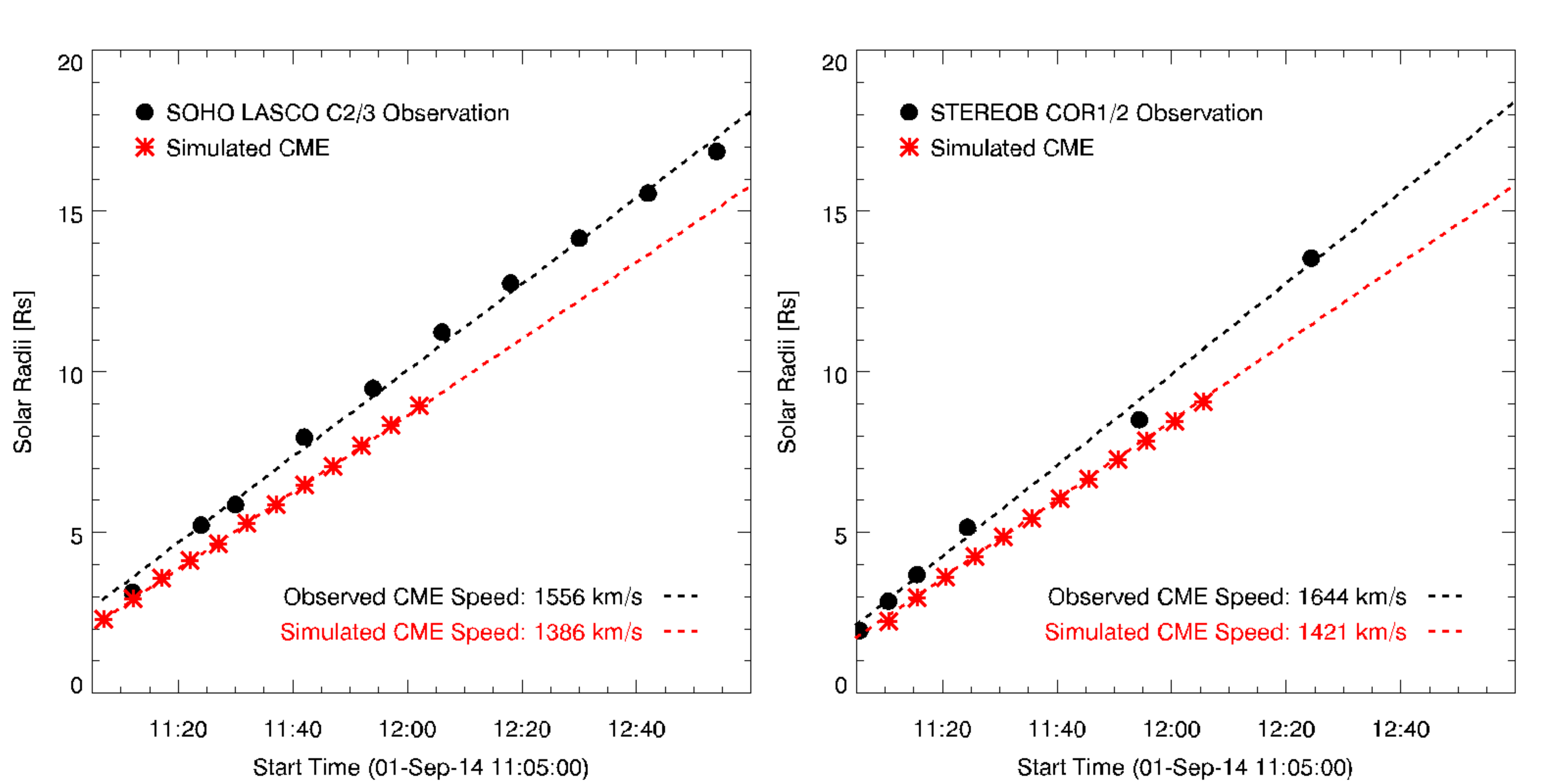}
\end{array}$
\end{center}
\caption{Comparison of CME height-time evolution in the observation (black dots; \emph{SOHO} LASCO C2/3 on the left and \emph{STEREO-B} COR1/2 on the right) and simulation from the synthesized white-light images (red stars). The simulated CME is about 10-14\% slower than the observed one.}
\label{fig:htplot}
\end{figure}

Since the CME propagation near the Sun is mainly observed by coronagraphs, we generate synthesized white-light images (Thomson-scattered white-light brightness) and compare them with observations. The top panel of Figure~\ref{fig:wl} shows the observations from \emph{SOHO}/LASCO C2 and \emph{STEREO-B}/COR1. The bottom panel shows the synthesized white-light images. The color scale shows the relative total brightness changes with respect to the pre-event level. At the time of the Sept14 event, \emph{STEREO-B} and \emph{SOHO} were separated by $\sim$161$^{\circ}$ therefore observing the Sun from nearly opposite directions. By comparing the observation and simulation from two different viewpoints, we find that the observed CME is adequately simulated in terms of the direction of propagation and the width. Note that the absolute brightness comparison between the observation and simulation requires advanced calibration of the observational data as well as the inclusion of the contribution from the F corona (light scattered by interplanetary dust) in the simulation data, which are beyond the scope of this study. The reader is referred to previous studies for such model validation \citep[e.g.,][]{chip08, jin17a}. Also, there is a distinct feature (marked in Figure~\ref{fig:wl}) around the CME leading edge in the LASCO C2 image, which might imply that the corresponding shock front can deviate from a typically circular or dome shape. We speculate that this feature might be related to the complex and dynamically changing
background solar wind environment (e.g., affected by previous CMEs or coronal disturbances), which is not captured by the current simulation. We further compare the observed and simulated CME speeds by tracking the height-time (HT) history of the CME leading edge, as shown in Figure~\ref{fig:htplot}. The black dots show measurements from \emph{SOHO}/LASCO C2/3 (left panel) and \emph{STEREO-B} COR1/2 (right panel), while the red asterisks show corresponding measurements from the synthesized white-light images. We use the moment when the observed and simulated CMEs are around the same height as a guidance to calibrate the start time of the simulation in terms of the real observation. With this assumption, the first appearance of CME in the LASCO C2 field of view (11:12:05 UT) corresponds to $t = 10$~minutes in the simulation. In general, the CME HT history is well reproduced in the simulation, with the simulated CME being slightly slower by about 10-14\%.

\subsection{Field Connectivity Evolution}
\label{subsect_connect}

In the course of the eruption, the flux rope interacts and reconnects with the magnetic fields of the source AR as well as the global coronal field. As a result, the magnetic field configuration and connectivity can change dramatically, which could significantly influence the transport of the accelerated particles. With this global MHD simulation of the Sept14 event, we now investigate the field connectivity evolution in detail during the first hour of CME evolution.

\begin{figure}[h]
\begin{center}$
\begin{array}{c}
\includegraphics[scale=0.9]{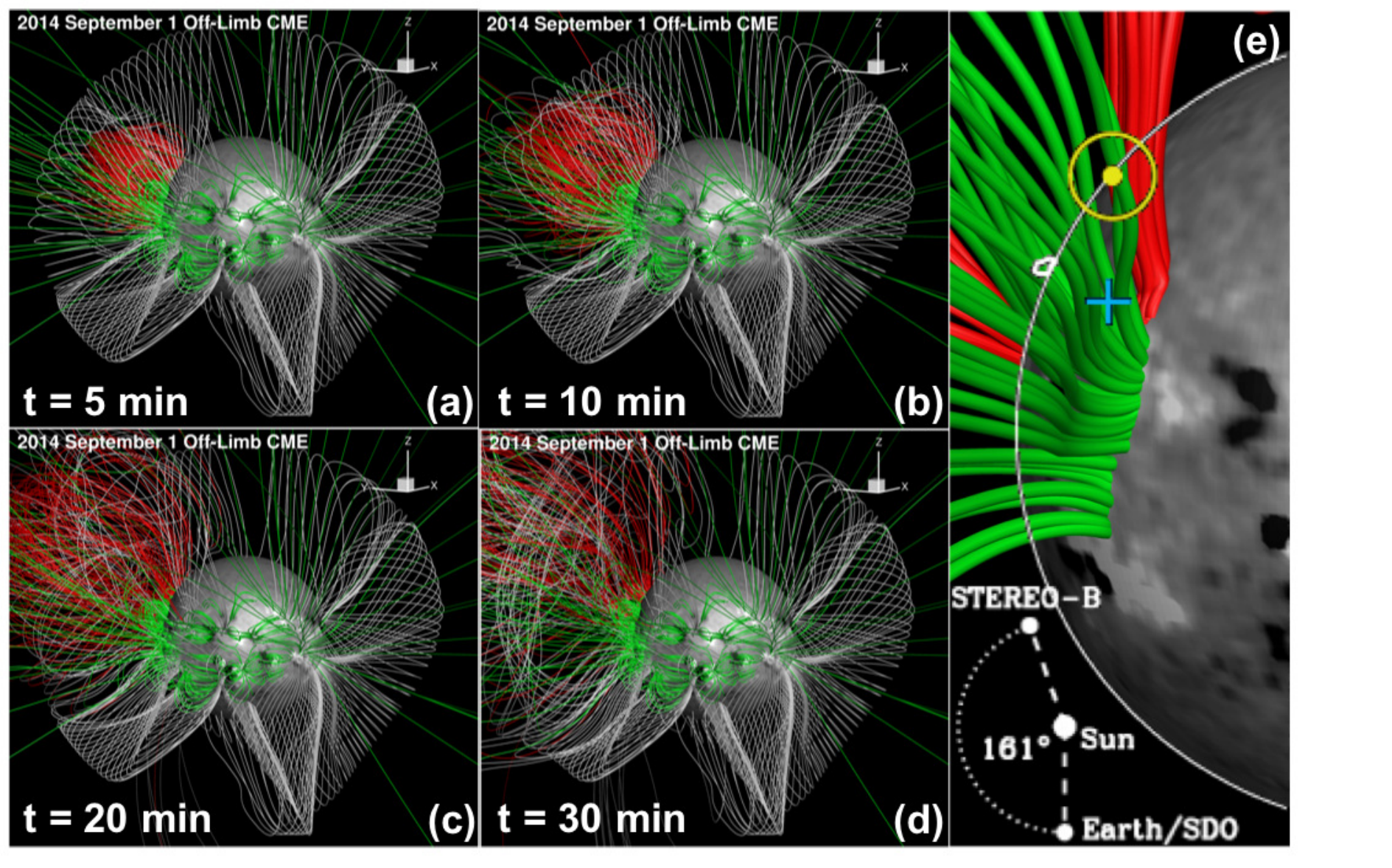}
\end{array}$
\end{center}
\caption{Magnetic field evolution in the first 30 minutes after the flux rope eruption. (a)-(d) show the 3D field configuration (viewed from the Earth) at $t$ = 5, 10, 20, and 30 minutes. The red field lines represent the flux rope. The white field lines represent the large-scale helmet streamers. The green field lines are selected surrounding active region as well as open field lines. (e) Selected field lines near the \emph{Fermi}-LAT $\gamma$-ray emission region from the simulation at $t$ = 30 minutes. The yellow dot and circle indicate the LAT $>$ 100 MeV emission centroid and 68\% error radius of 100$\arcsec$, respectively. The white contour shows the 6-12 keV \emph{RHESSI} source. The blue plus represents the projected BTL position of the \emph{STEREO} flare ribbon centroid. The green field lines are connecting to the CME-driven shock and the red field lines to the flare/CME source region behind the limb.}
\label{fig:3d_evo}
\end{figure}

Figures~\ref{fig:3d_evo}a-d show the 3D magnetic field configurations at selected times (5, 10, 20, and 30 minutes). Magnetic reconnection between the erupting flux rope (red) and the surrounding field lines (green and white) is evident, especially after the first 10 minutes. The interaction between the flux rope and the large-scale helmet streamers significantly changes the global corona configuration around the CME source region. The helmet streamers are opened up by reconnection or stretched by the CME expansion. Specifically, we further examine the field line connectivity around the \emph{Fermi} $\gamma$-ray emission region at $t = 30$~minutes (shown in Figure~\ref{fig:3d_evo}e). The derived \emph{Fermi} $\gamma$-ray emission centroid and 68\% uncertainty circle (adapted from \citealt{ackermann17}) are overlaid on the simulation data. The green field lines are the pre-existing open field connected to the CME-driven shock after $t \sim 6$~minutes. The red field lines are the closed field connected to the flaring AR. These field lines were not present before, but started to develop $\sim$5~minutes after the eruption through magnetic reconnection between the flux-rope magnetic field and the global coronal field.

\begin{figure}[h]
\begin{center}$
\begin{array}{c}
\includegraphics[scale=1.0]{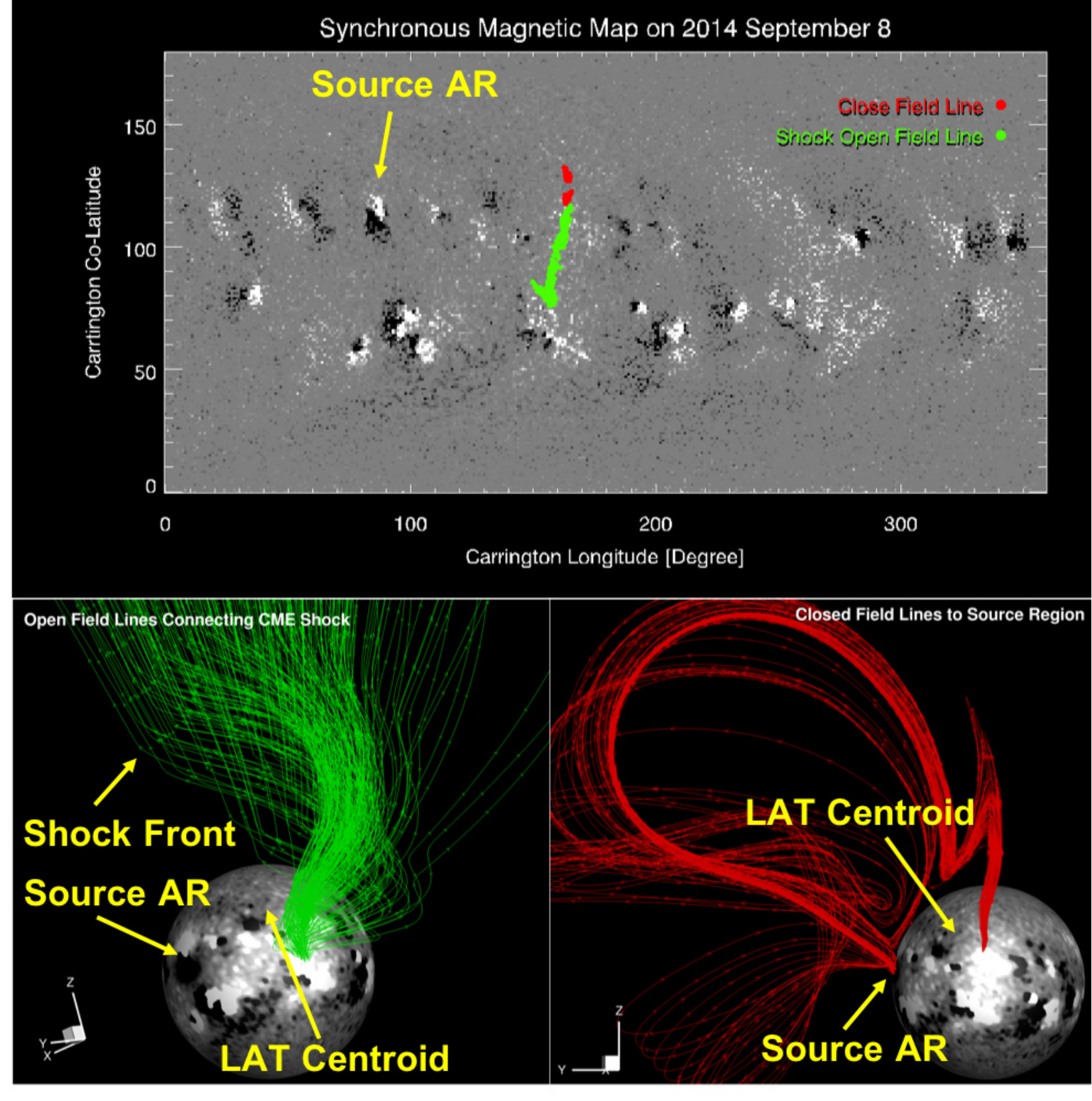}
\end{array}$
\end{center}
\caption{Top panel: Footpoints of the field lines connecting to the CME-driven shock (open field, green) and the source region (closed field, red) on the magnetic map. Bottom panels: the 3D field configuration at $t$ = 30 minutes for the open (left) and closed (right) field lines seen from different view points. Several important features (e.g., source active region, Fermi/LAT centroid, shock front) are marked. Note that in the lower panels, we use a smaller saturation threshold for the magnetogram (than in the upper panel) in order to display weaker magnetic fields than the ARs.}
\label{fig:open_close}
\end{figure}

To investigate the details of these two types of field lines, we further mark their photospheric footpoints on the magnetic field map in the top panel of Figure \ref{fig:open_close}. The closed field line regions (red) are relatively compact, compared with the elongated open field line region (green) to the south. The bottom panel of Figure \ref{fig:open_close} shows the 3D configuration of these two types of field lines. As evident in this plot, the open field lines change directions abruptly due to the rapid expansion the CME and CME-driven shock. For the closed field lines, the configuration is more complex with twisted large-scale loops. It appears that some of these field lines result from reconnection between the erupting flux rope and the nearby helmet streamers.

\subsection{CME-driven Shock Evolution}
\label{subsect_shock}

After the eruption, the flux rope drives a shock in the corona that propagates freely into the heliosphere. CME-driven shocks are believed to be responsible for acceleration of particles through the diffusive shock acceleration (DSA) mechanism (e.g., \citealt{axford77}) that produces the so-called gradual solar energetic particle (SEP) events \citep{reames99}. Due to the nonuniform background environment, the CME-driven shock evolution is highly spatially dependent. For example, a shock that is propagating into the fast solar wind could acquire a higher shock speed therefore leading to a higher stand-off distance from the flux rope driver \citep{jin17a}. Also, the shock parameters could vary significantly over the shock front, which can significantly affect the acceleration process \citep{chip05, li12}. Based on the white-light observations from SOHO and STEREO, several methods have been developed to derive the shock parameters directly from the data (e.g., \citealt{rouillard16, lario17, kwon18}). In this study, with the data-driven MHD simulation of the Sept14 event, we can track the shock location and key parameters (e.g., the compression ratio, shock Alfv\'{e}n Mach number, shock speed, and shock obliquity angle $\theta_{Bn}$) during the CME evolution. The shock obliquity angle $\theta_{Bn}$ refers to the angle between the shock normal (see equation~[\ref{eq_shock}]) and the upstream magnetic field. As shown below, such analysis can provide a more comprehensive picture of the shock as to its configuration and properties over the area linking back to the visible side of the Sun, where LAT $\gamma$-rays were detected.

We first determine the shock location at each time step by using the proton temperature gradient criteria \citep{jin13}. At each shock location, the shock normal is determined by using the magnetic coplanarity ${\bf (B_{d}-B_{u})}\cdot {\bf n}=0$ \citep{abraham72, lepping71}:

\begin{equation}
{\bf n}=\pm\frac{({\bf B_{d}}\times{\bf B_{u}})\times({\bf B_{d}}-{\bf B_{u}})}{|({\bf B_{d}}\times{\bf B_{u}})\times({\bf B_{d}}-{\bf B_{u}})|}
\label{eq_shock}
\end{equation}
where $\bf B_{d}$ and $\bf B_{u}$ represent downstream and upstream magnetic field respectively. Note that this method fails for $\theta_{Bn} = 0^{\circ}$ or $90^{\circ}$, which we found to be very rare in the actual simulations. The $\pm$ sign is determined by assuming a forward moving shock in the heliocentric coordinate. We then determine the upstream/downstream plasma parameters, from which the shock parameters (e.g., the compression ratio, shock speed, shock Alfv\'{e}n Mach number, and shock obliquity angle) can be calculated accordingly.

\begin{figure}[h]
\begin{center}$
\begin{array}{c}
\includegraphics[scale=0.85]{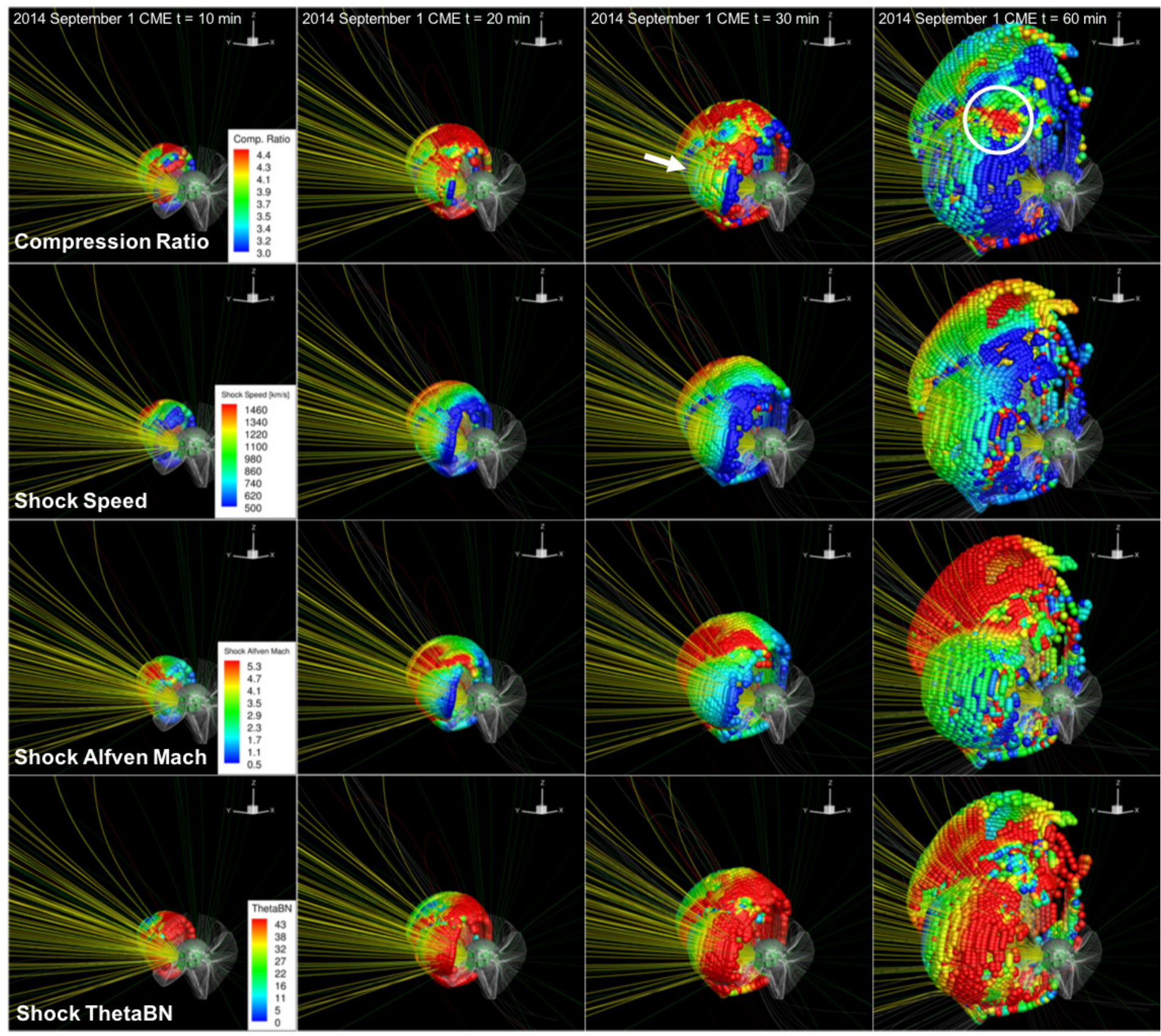}
\end{array}$
\end{center}
\caption{Evolution of shock parameters at $t$ = 10, 20, 30, and 60 minutes from left to right. The top to bottom panels represent the compression ratio, shock speed, shock Alfv\'{e}n Mach number, and shock $\theta_{Bn}$. The yellow field lines represent the open field near the \emph{Fermi}-LAT $\gamma$-ray emission region connected to the CME-driven shock. The white arrow points to the shock surface connected back to the visible side of the Sun. The white circle in the upper right panel marks the possible shock-shock interaction region (see text).}
\label{fig:shock_evo}
\end{figure}

Figure \ref{fig:shock_evo} shows the shock geometry evolution during the first hour of the simulation. The color scales on the shock surface represent the four shock parameters (from top to bottom): the compression ratio, shock speed, shock Alfv\'{e}n Mach number, and shock $\theta_{Bn}$. The yellow field lines represent the open field near the \emph{Fermi} emission region (shown in Figures \ref{fig:3d_evo} and \ref{fig:open_close}). Based on the simulation, we found that the CME-driven shock started to intersect the open field lines around $t = 6$~minutes, when the fastest part of the shock reached $\sim$3 R$_{\odot}$. This finding is consistent with the estimation of the CME located at $\sim$2.5 R$_{\odot}$ at the onset of the \emph{Fermi}-LAT emission \citep{ackermann17}. However, we should note that the part of the shock intersecting the open field is closer to the Sun at $\sim$1.6 R$_{\odot}$. At $t = 20$~minutes, the CME-driven shock covered the entire open-field region around $\sim$3 R$_{\odot}$ linking to the front side of the Sun. Furthermore, the derived $\theta_{Bn}$ suggests that this part of the shock is a quasi-perpendicular shock with a mean $\theta_{Bn}\sim73^{\circ}$. Another observational fact worth mentioning is the EUV wave observed in this event. The EUV wave from the source region arrived at the open field region (connecting to the CME-driven shock) by 11:20 UT as shown in online movies (http://aia.lmsal.com/AIA\_Waves), an extension of the \citet{nitta13} study. There is a possibility that this EUV wave may trace the low-corona flank of the shock, as see in several other eruptive events (e.g., \citealt{carley13}), including the recent X8.2 flare on 2017 September 10 \citep{gopal18a, liu18, morosan18}, which was the first (and the only one so far) long-duration \emph{Fermi} flare associated with a ground level enhancement event \citep{omodei18}.  

At $t = 30$~minutes, the shock surface starts to deviate from its initial spherical shape due to the non-uniform background solar wind condition. In particular, one part of the shock (as marked by a white arrow in Figure \ref{fig:shock_evo}), with open field lines crossing it, propagates into a fast wind region originating from an on-disk coronal hole and acquires a higher speed. This process may also lead to a ``shock-shock" interaction at the boundary between the two shock surfaces that causes an elevated shock compression ratio and Alfv\'{e}n Mach number at $t = 60$~minutes (marked with a white circle in the upper-right panel of Figure \ref{fig:shock_evo}). Note that the open field lines connected to this shock interaction region are closer to the \emph{Fermi} $\gamma$-ray emission region. Since the compression ratio, shock Alfv\'{e}n Mach number, and shock geometry are key parameters for DSA of SEPs, this part of the shock can be favorable for accelerating particles to higher energies \citep{chip05, sokolov06, li12, zhao14, hu17, hu18}.

\begin{figure}[h]
\begin{center}$
\begin{array}{c}
\includegraphics[scale=0.43]{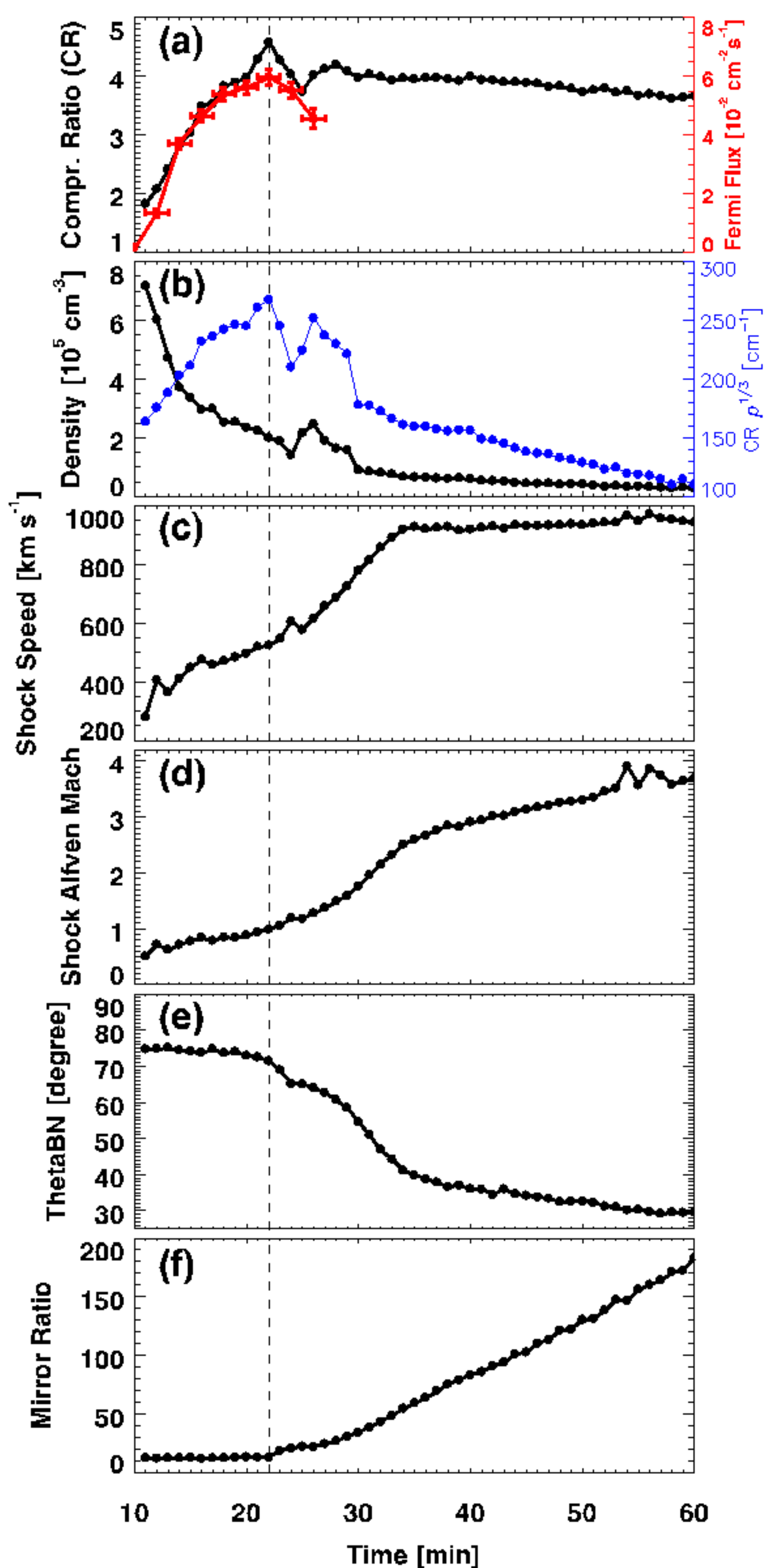}
\end{array}$
\end{center}
\caption{Evolution of shock parameters averaged over the shock surface connected back to the visible side of the Sun for (a) Compression ratio (with \emph{Fermi}-LAT $>$100 MeV flux overlaid, red, right axis); (b) upstream local plasma number density (with an empirical quantity $CR\cdot\rho^{1/3}$ combining the compression ratio CR and density $\rho$ overlaid, blue, right axis); (c) shock speed; (d) shock Alfv\'{e}n Mach number;  (e) shock obliquity angle $\theta_{Bn}$; and (f) magnetic mirror ratio $\eta$ derived from the simulation. The vertical dashed line marks the peak time of the compression ratio, co-temporal with the peak of the \emph{Fermi} $\gamma$-ray flux.}
\label{fig:shock_profile}
\end{figure}

We obtain the shock parameters averaged over the portion of the shock surface that is connected back to the visible side of the Sun. The temporal evolution of the resulting four key shock parameters as well as the average upstream local plasma density is shown in Figure \ref{fig:shock_profile} and described as follows:

\begin{enumerate}

\item
The shock compression ratio (Figure \ref{fig:shock_profile}a) increases rapidly from $\sim$1.8 at $t \sim$10~minutes to $\sim$4.6 at $t \sim 20$~minutes\footnote{The maximum compression ratio is slightly larger than 4 (strong shock limit) due to the non-ideal processes (e.g., heat conduction, other compression effects) or merged background density gradient in the simulation.}, and then gradually decreases to $\sim$3.7 at $t = 60$~minutes. This evolution trend closely follows that of the \emph{Fermi}/LAT $\gamma$-ray flux profile (red curve), with a similar $\sim$10~minute duration of the rapid rise phase.

\item
The average local plasma density at the shock front (Figure \ref{fig:shock_profile}b) is another important parameter, which is related to the seed population and the number of particles available for shock acceleration. We also plot an empirical quantity $CR\cdot\rho^{1/3}$ (blue curve), a product of the shock compression ratio (CR) and the ambient density to a $1/3$ power (heuristically selected to match the temporal trend of the Fermi $\gamma$-ray flux). The density generally decreases with time as the CME travels away from the Sun, which causes this empirical quantity to decrease after its initial increase during the first ~20 minutes. This could potentially explain the simultaneous decrease in the Fermi $\gamma$-ray flux, even though the shock compression ratio and Mach number remain high.

\item
The shock speed (Figure \ref{fig:shock_profile}c) shows a gradual increase from $\sim$400~km~s$^{-1}$ at $t \sim$10~minutes to $\sim$1000~km~s$^{-1}$ at $t \sim$35~minutes and then remains roughly constant.

\item
Likewise, the shock Alfv\'{e}n Mach number (Figure \ref{fig:shock_profile}d) gradually increases from $\sim$1 to $\sim$3 during the $t \sim$10-35~minutes interval and then grows even more slowly to $\sim$4 at $t \sim$60~minutes.

\item
The shock obliquity angle analysis (Figure \ref{fig:shock_profile}e) shows that the shock is originally a quasi-perpendicular shock before $t \sim$30~minutes (with $\theta_{Bn} \sim 75 ^\circ$ at $t \sim$10~minutes) and evolves into a quasi-parallel shock (with $\theta_{Bn} \sim 30 ^\circ$ at $t = 60$~minutes). The most rapid decrease in $\theta_{Bn}$ occurs during $t \sim$22-35~minutes, the onset of which coincides with the peak time ($t=22$~minutes) of the compression ratio and \emph{Fermi} $\gamma$-ray flux shown in Figure \ref{fig:shock_profile}a. Note that the same trend of shock obliquity angle variation during CME evolution was also found by \citet{chip05}.

\end{enumerate}

\section{Discussions}
\label{sect_discuss}

\subsection{Magnetic Connectivity: CME-Shock vs. Flare Site}
\label{subsect_magconnect}
As noted earlier in Section~\ref{subsect_connect}, the \emph{Fermi} $\gamma$-ray emission centroid is closer to the footpoints of the closed field lines connecting the source AR than those of the open field lines connecting the CME shock. It is possible that the particles accelerated at the flaring site or the areas passed by the coronal shock \citep{hudson18} can be trapped in the closed field through some mechanisms (e.g., \citealt{sheeley04}), re-accelerated, and then transported to the front side of the Sun through the connectivity established by the interaction between the erupting flux rope and global corona field. However, we would like to emphasize that, based on the present simulation result, it is difficult to unambiguously distinguish the potential contributions from the two groups of field lines to the observed LAT emission for four reasons. (i) The northern portion of the open-field footpoints is adjacent \emph{in space} to both the closed-field footpoints and the LAT centroid; (ii) The appearances of the closed field ($t \sim 5$~minutes) and open field ($t \sim 6$~minutes) are very close \emph{in time};	(iii) The current localization of the LAT centroid is based on an assumption of a point source for the $\gamma$-ray emission and can change if the actual source shape deviates from this assumption; (iv) Because of the proximity between the open and closed field lines, cross-field diffusion (e.g., \citealt{zhang03}) can allow shock-accelerated energetic particles to access the closed field lines as well.

\subsection{The CME-shock and $\gamma$-ray Connection}
\label{subsect_shockconnect}

However, our simulation result of the Sept14 event together with our initial inspection of other \f BTL flares, does reveal some unique and attractive features about the CME-driven shock linked to the observed $\gamma$-rays:

\begin{enumerate}

\item
Since the shock compression ratio is one of the key parameters in the DSA mechanism that determines the energetic particle production at the shock (e.g., the particle spectral index), the temporal correlation noted in Item~1 (Section~\ref{subsect_shock}) above indicates an intimate relation between the $\gamma$-ray flux and the shock particle production. This provides clear evidence supporting the mechanism that (at least some of) the $\gamma$-ray producing particles are accelerated by the CME-driven shock.

\item
The open field is connected to a quasi-perpendicular shock early on (Item~4 in Section~\ref{subsect_shock}), which is generally believed to be an efficient particle accelerator if the upstream coronal or heliospheric magnetic field is sufficiently turbulent (e.g.,\citealt{giacalone05, tylka05}). Furthermore, a recent \emph{in-situ} observation of Saturn's bow shock from the Cassini spacecraft \citep{masters17} shows that energetic electrons were only detected \emph{downstream} of the quasi-perpendicular shock, which suggests the potential importance of a quasi-perpendicular shock in accelerating particles that could escape the \emph{downstream} and propagate \emph{back to the Sun} to produce $\gamma$-rays.

\item
Another piece of supporting evidence for shock acceleration is that in all three identified \emph{Fermi} BTL events (Sept14, {\it SOL2013-10-11}, and {\it SOL2014-01-06}; \citealt{pesce15, ackermann17}), there are pre-existing open field lines (e.g., in on-disk coronal holes) near the $\gamma$-ray emission region, which could be potentially connected to the CME-driven shock.

\end{enumerate}

By using a 3D triangulation technique, \citet{plotnikov17} reconstructed the CME-driven shock structure from white-light observations of the Sept14 event. With the density and magnetic field information obtained from a \emph{static} solar corona constructed with the Magnetohydrodynamic Algorithm outside a Sphere (MAS) model \citep{lionello09}, the time-dependent distribution of the shock Mach number and obliquity angle were approximately derived. They found that the Mach number shows a rapid increase to supercritical values after the type-II burst onset and the shock has a quasi-perpendicular geometry during the $\gamma$-ray emission, which are in general agreement with our results. However, an important distinction between their and our studies is that, instead of using a \emph{static} coronal model, we self-consistently simulated the \emph{dynamic} evolution of the CME and the CME-driven shock. This allows us to track the detailed temporal evolution of the shock and derive the shock compression ratio, which is of critical importance to particle acceleration by shocks. In addition, we found that the shock geometry evolves and changes from quasi-perpendicular to quasi-parallel, instead of remaining quasi-perpendicular all the time.

We also briefly discuss the shock Alfv\'{e}n Mach number evolution derived from the simulation. Note that when this number is around unity (see Item~3 in Section~\ref{subsect_shock}), stochastic acceleration of particles by plasma turbulence (e.g., in the downstream of the shock) is more efficient than DSA \citep{petrosian16}. Stochastically accelerated particles could also serve as the seed population to be further accelerated by the shock. This could be the case early on, when the shock compression ratio is also relatively low, and could be related to the rapid rise in the detected LAT $\gamma$-ray flux. Later on, when the shock Alfv\'{e}n Mach number and compression ratio are sufficiently large, shock acceleration would be more important and can account for the gradual, long-duration $\gamma$-ray emission. Therefore, it is likely that, in addition to DSA, stochastic acceleration could also play a role in the Sept14 event, especially in the early stage. On the other hand, as mentioned in \S 3.1, the simulated CME speed is 10-14\% slower than the observed one. Considering the higher shock speed in the simulation, the Alfv\'{e}n Mach number will also be higher. The critical Alfv\'{e}n Mach number is estimated between 1 and 1.7 for the solar wind plasma with $\gamma = \frac{5}{3}$ and $\beta = 1.0$ \citep{edmiston84}. Therefore, it is also possible that the shock in the simulation already becomes supercritical in the early stage. In summary, we believe multiple acceleration mechanisms could be important in the beginning but DSA might be the dominant one after 20 minutes.

\subsection{Effect of Magnetic Mirroring}
\label{subsect_magmirror}

Finally, we discuss the possibility of particle transport back to the Sun from the CME-driven shock. This may appear difficult because of strong magnetic mirroring due to the high degree of convergence of magnetic field lines  toward the Sun (with a mirror ratio of $\eta=B_{\odot}/B _{\rm CME}\gg 1$) and thus extremely small loss cones (e.g., a few degress; \citealt{klein18}, see their Section 8.4.4). In an ideal \emph{scattering-free} environment, this could potentially prevent particles from reaching the photosphere to produce $\gamma$-rays. In reality, however, the CME environment in general, and the shock downstream region in particular, are most likely \emph{highly turbulent}, rendering a sufficiently short scattering mean free path which can continuously scatter particles into the loss cone, thus precipitating to the Sun. How fast and what fraction of the particles can reach the photosphere depend on the relative values of the duration $\Delta T$ of the emission and the escape time from the trap, $T_{\rm esc}$. The latter depends on the mirror ratio $\eta$, the scattering time $\tau_{\rm sc}$, and the crossing time from the shock back to the Sun, $\tau_{\rm cross}\sim L/v_p$, where $v_p\sim c$ is the velocity of GeV protons responsible for $\gamma$-rays and $L=v_{\rm CME}\Delta T$ is the distance between the CME and the Sun. With some analytical and numerical treatments, \citet{malyshkin01} gave an approximate relationship between the escape time and the three variables $\eta$, $\tau_{\rm sc}$, and $\tau_{\rm cross}$ (see Figure 2, Petrosian 2016):
\begin{equation}
T_{\rm esc}=\tau_{\rm cross}\left(2\eta+\frac{\tau_{\rm cross}}{\tau_{\rm sc}}+\ln\eta\frac{\tau_{\rm sc}}{\tau_{\rm cross}}\right).
\end{equation}
Recent numerical simulations by \citet{effenberger18} have confirmed this relation. The upshot of this result is that for an \emph{isotropic} pitch-angle distribution, $T_{\rm esc}\sim 2\eta \tau_{\rm cross}$ for $\tau_{\rm sc}\sim \tau_{\rm cross}$ and $\eta \gg 1$, which means that $T_{\rm esc}/\Delta T\sim 2\eta v_{\rm CME}/c \sim \eta/100$ for a CME speed of $v_{\rm CME}=1,500$ km/s. Therefore, for $\eta\lesssim 100$, we have $T_{\rm esc} \lesssim \Delta T$; i.e., a large fraction of the downstream GeV protons can reach the photosphere within the emission duration $\Delta T$. We have tracked the history of the magnetic field strength along the field lines connecting the shock to the solar surface in our simulation and found that the median value of the mirror ratio $\eta$ increases from $\sim 10$ to $\sim 100$ during the first hour (Figure \ref{fig:shock_profile}f). Thus, for the first few hours, which is the duration of most Fermi events, the difficulty of particle transport back to the Sun due to magnetic mirroring can be overcome with a scattering mean free path of order of a solar radius or a scattering time of $\tau_{\rm sc}\sim 2$ s. For protons with gyro-frequency of $\Omega_p\sim15\cdot\left[\frac{B}{mG}\right]$  Hz, and magnetic field of $\sim$100 mG (the typical average value on the shock surface in the first hour of simulation), this would require a fractional turbulence energy ($\delta B/B)^2\sim 1/(\Omega_p \tau_{\rm sc})\sim 3\times10^{-4}$. According to \citet{effenberger18}, the situation is similar for a \emph{pancake} pitch-angle distribution. But for a distribution \emph{beamed} along the field lines, the escape time will be shorter and thus facilitate particle precipitation back to the Sun.

\section{Summary \& Conclusion}
\label{sect_summary}

In this study, we simulated the CME associated with a well-known \emph{Fermi} BTL flare on September 1, 2014 by using a data-driven global MHD model AWSoM within SWMF. We tracked the dynamic evolution of the global magnetic field and the CME-driven shock and investigated the magnetic connectivity between the shock and the region around the centroid of the \emph{Fermi}-LAT $\gamma$-ray source. We found supporting evidence for the hypothesis that the observed $\gamma$-ray emission is produced by particles that are accelerated in the CME environment and escape the shock downstream region along magnetic field lines connected to regions on the Sun far away from the hosting AR of the flare. Our specific findings are summarized as follows:

\begin{enumerate}

\item
To enable the high-energy particle precipitation and thus $\gamma$-ray emission on the front side of the Sun, certain \emph{magnetic connectivity} must be established between the emission region and the flare source AR or the CME-driven shock. In our simulation, both types of connections are present and appear close in space and time within the first few minutes of the event, as a result of the interaction between the erupting flux-rope magnetic field and the global solar corona. The CME-driven shock is connected to the front side of the Sun by open magnetic field lines that originate from an on-disk coronal hole. This part of the shock surface is away from the flux-rope driver and the shock nose that are propagating in a different direction. Such open-field configurations represent a favorable condition for connecting the CME-driven shock back to the solar surface, and have been identified in all three \emph{Fermi} BTL events reported so far.

\item
Within the shock surface connected to the front side of the Sun, the \emph{shock properties} vary significantly with time and space. The temporal evolution of the compression ratio and thus the rate of particle acceleration by the shock are closely correlated with the \f $\gamma$-ray flux, suggestive of a causal relationship. In addition, this part of the shock is initially a quasi-perpendicular shock and later evolves to a quasi-parallel shock, the former of which is believed to be an effective particle acclerator.

\item
These findings provide strong support for the aforementioned hypothesis and indicate that the \emph{CME-driven shock} can play an important role in accelerating particles that then travel back to the Sun to produce observed $\gamma$-rays. In addition to DSA, \emph{stochastic acceleration} by plasma turbulence may play a role as well, especially in the shock downstream region and during the early stage of the event.

\end{enumerate}

The present study is among the first attempts to solve the puzzle of \f BTL $\gamma$-ray flares. The identified mechanisms, in general, could be at work in on-disk \emph{Fermi} flares as well and can potentially solve another puzzle, i.e., long-duration $\gamma$-ray flares. BTL and long-duration $\gamma$-ray flares could be viewed as two faces of the same puzzle, with the $\gamma$-ray emission being \emph{spatially separated} in the former and \emph{temporally delayed} in the latter from the main flare emission commonly observed at longer wavelengths. In fact, recent statistical studies show that long-duration $\gamma$-ray flares observed by \emph{Fermi}/LAT always associated with wide, fast CMEs \citep{winter18} and Type II radio bursts \citep{gopal18b}. These results strongly support that long-duration $\gamma$-rays are produced by shock-accelerated protons precipitating back to the Sun.

Ultimately, one needs to self-consistently couple MHD simulations with particle acceleration, escape, and transport models (e.g., Borovikov et al. 2017, \citealt{hu18}). Furthermore, a comparative study is needed among not only the BTL events but also the on-disk events, by combining observational and simulation efforts, which we plan to pursue in future studies.

\begin{acknowledgements}
We are very grateful to the referee for invaluable comments that helped improve the paper. M.J. and W.L. were supported by NASA's {\it SDO}/AIA contract NNG04EA00C to LMSAL. W.L. by NASA HGI grants NNX15AR15G and NNX16AF78G and LWS grant NNX14AJ49G. N.V.N by NSF grant AGS-1259549. F.E. by NASA grant NNX17AK25G. G.L. by NASA grants NNX17AI17G and NNX17AK25G. W.M. by NASA grant NNX16AL12G and NSF AGS 1322543. We are thankful for the use of the NASA Supercomputer Pleiades at Ames and to its supporting staff for making it possible to perform the simulations presented in this paper.

The \emph{Fermi} LAT Collaboration acknowledges generous ongoing support from a number of agencies and institutes that have supported both the development and the operation of the LAT as well as scientific data analysis. These include the National Aeronautics and Space Administration and the Department of Energy in the United States, the Commissariat \'{a} l'Energie Atomique and the Centre National de la Recherche Scientifique / Institut National de Physique Nucl\'{e}aire et de Physique des Particules in France, the Agenzia Spaziale Italiana and the Istituto Nazionale di Fisica Nucleare in Italy, the Ministry of Education, Culture, Sports, Science and Technology (MEXT), High Energy Accelerator Research Organization (KEK) and Japan Aerospace Exploration Agency (JAXA) in Japan, and the K. A. Wallenberg Foundation, the Swedish Research Council and the Swedish National Space Board in Sweden. {\it SDO} is the first mission of the NASA's Living With a Star (LWS) Program.
\end{acknowledgements}

\newpage

\newpage

\newpage

\newpage

\newpage

\newpage

\newpage

\newpage

\end{document}